\documentclass[iop,apjl]{emulateapj}

\usepackage{graphicx}  
\usepackage{dcolumn}   
\usepackage{bm}        
\usepackage{amsfonts,amsmath,amssymb,mathrsfs}
\usepackage{color}

\shorttitle{AIC of NS to BH}

\begin{document}

\title{General Relativistic Simulations of Accretion Induced Collapse
  of Neutron Stars to Black Holes}

\author{Bruno {Giacomazzo}\altaffilmark{1} and Rosalba
  {Perna}\altaffilmark{2}}

\altaffiltext{1}{JILA, University of Colorado and National Institute
  of Standards and Technology, Boulder, CO 80309, USA}

\altaffiltext{2}{JILA and Department of Astrophysical and Planetary
  Sciences, University of Colorado, Boulder, CO 80309, USA}

\begin{abstract}

Neutron stars (NSs) in the astrophysical universe are often surrounded
by accretion disks. Accretion of matter onto an NS may increase its
mass above the maximum value allowed by its equation of state,
inducing its collapse to a black hole (BH). Here we study this process
for the first time, in three-dimensions, and in full general
relativity. By considering three initial NS configurations, each with
and without a surrounding disk (of mass $\sim7\%~M_{\rm NS}$), we
investigate the effect of the accretion disk on the dynamics of the
collapse and its imprint on both the gravitational wave (GW) and
electromagnetic (EM) signals that can be emitted by these sources. We
show in particular that, even if the GW signal is similar for the
accretion induced collapse (AIC) and the collapse of an NS in vacuum
(and detectable only for Galactic sources), the EM counterpart could
allow us to discriminate between these two types of events.  In fact,
our simulations show that, while the collapse of an NS in vacuum
leaves no appreciable baryonic matter outside the event horizon, an
AIC is followed by a phase of rapid accretion of the surviving disk
onto the newly formed BH. The post-collapse accretion rates, on the
order of $\sim 10^{-2}~M_{\odot} ~{\rm s}^{-1}$, make these events
tantalizing candidates as engines of short gamma-ray bursts.

\end{abstract}

\keywords{accretion, accretion disks --- gamma-ray burst: general
  ---gravitational waves --- methods: numerical --- stars: neutron}

\section{Introduction}

In the astrophysical universe, neutron stars (NSs) are often
surrounded by an accretion disk. Several observations of X-ray
binaries have indicated the presence of NSs in binary systems in which
the NS accretes from the less-compact companion
(e.g., \citealt{bildsten97}). Furthermore, isolated NSs are likely to
be surrounded by a fallback disk remnant of the supernova explosion
(e.g., \citealt{chev}), while numerical simulations have shown that the
accretion induced collapse (AIC) of a white dwarf may lead to an NS
surrounded by a disk with mass up to $\sim 0.8
M_{\odot}$~\citep{2010PhRvD..81d4012A}.

Even if the NS has initially a mass below the maximum limit, it can
soon become unstable to gravitational collapse by accreting matter
from the surrounding disk. Such AIC to a spinning black hole (BH)
could be a powerful source of gravitational waves (GWs), especially if
the accreting matter excites non-axisymmetric modes in the NS before
collapse. To date, no studies of such sources have been reported in
full general relativity. On the other hand, the collapse of NSs (in
vacuum) to BHs has been the subject of several investigations,
including the accurate extraction of the GW signals emitted by these
sources~\citep{1985PhRvL..55..891S, 2000PhRvD..61d4012S,
  2003PhRvD..67b4033S, 2005PhRvD..71b4035B, 2005PhRvL..94m1101B,
  2006PhRvL..97n1101B, 2007CQGra..24S.187B, 2011PhRvD..84b4022G}. All
these simulations have shown that the collapse is essentially
axisymmetric and that no torus is left behind after the creation of a
spinning BH. Because of this, the GW signals emitted by these
sources will be rather weak and detectable only by advanced LIGO and
Virgo if the sources were located in our Galaxy. Third generation
detectors, such as the planned Einstein Telescope, could instead
detect some of these sources up to a distance of $\sim 1$
Mpc~\citep{2011PhRvD..84b4022G}.

The AIC of an NS to a BH has also been suggested to be behind the central
engine of short gamma-ray bursts (GRBs, e.g., see
\citealt{2005astro.ph.10192M} and \citealt{2006ApJ...643L..13D}). The
idea is that an NS in a low-mass binary system could accrete enough
mass from the companion star to produce a spinning BH surrounded by an
accretion disk massive enough to power a relativistic jet and hence a
short GRB. 

Here we present the first numerical simulations describing the AIC of
an NS to BH via fully three-dimensional general relativistic
simulations. We follow the collapse of three initial NS
configurations, each with and without an accreting torus, and study
the impact of the accreting material on the dynamics of the collapse
and its imprint on the GW signal, as well as the post-collapse
evolution of the disk. Given that this is the first of such studies,
we consider non-magnetized NSs. We also note that, while we do not
expect magnetic fields to have a strong impact during the collapse
(because, even when considering highly magnetized NSs, the ratio of
magnetic to gas pressure is much lower than one), they can have an
important role after BH formation since they may extract energy from
the system and power relativistic jets~\citep{rez11}.

\begin{table*}[t!]
\begin{center}
\caption{Properties of the Initial Models\label{tab:id}}
\begin{tabular}{lcccccccccccc}
\tableline\tableline
Model &
\multicolumn{1}{c}{$\rho_c$} &
\multicolumn{1}{c}{$r_p/r_e$} &
\multicolumn{1}{c}{$M_{b}$} &
\multicolumn{1}{c}{$M$} &
\multicolumn{1}{c}{$R_e$} &
\multicolumn{1}{c}{$J/M^2$} &
\multicolumn{1}{c}{$C$} &
\multicolumn{1}{c}{$l_0$} &
\multicolumn{1}{c}{$\delta$} &
\multicolumn{1}{c}{$R_{\mathrm{in}}$} &
\multicolumn{1}{c}{$R_{\mathrm{out}}$} &
\multicolumn{1}{c}{$M_{\mathrm{torus}}$}
\\
&
\multicolumn{1}{c}{$(10^{15}\, \mathrm{g\,cm^{-3}})$}&
&
\multicolumn{1}{c}{$(M_{\odot})$} &
\multicolumn{1}{c}{$(M_{\odot})$} &
\multicolumn{1}{c}{(km)} &
&
&
\multicolumn{1}{c}{$$} &
\multicolumn{1}{c}{$$} &
\multicolumn{1}{c}{(km)} &
\multicolumn{1}{c}{(km)} &
\multicolumn{1}{c}{$(M_{\odot})$}
\\
\tableline
1a   & $1.91$ & $0.85$ & $1.90$ & $1.73$ & $12.2$ & $0.36$ & $0.21$ & ------ & ------  & ------  & ------  & ------  \\
1b  & $1.91$ & $0.85$ & $1.90$ & $1.73$ & $12.2$ & $0.36$ & $0.21$ & $3.63$ & $-0.60$ & $13.6$  & $31.6$  & $0.141$ \\
2a   & $1.85$ & $0.75$ & $1.98$ & $1.80$ & $13.2$ & $0.47$ & $0.20$ & ------ & ------  & ------  & ------  & ------  \\
2b  & $1.85$ & $0.75$ & $1.98$ & $1.80$ & $13.2$ & $0.47$ & $0.20$ & $3.61$ & $-0.74$ & $14.7$  & $32.5$  & $0.143$ \\
3a   & $1.85$ & $0.65$ & $2.05$ & $1.86$ & $14.4$ & $0.55$ & $0.19$ & ------ & ------  & ------  & ------  & ------  \\
3b  & $1.85$ & $0.65$ & $2.05$ & $1.86$ & $14.4$ & $0.55$ & $0.19$ & $3.60$ & $-0.80$ & $15.5$  & $33.4$  & $0.145$ \\
\tableline
\end{tabular}
\tablecomments{The columns refer respectively to the name of the
  model, the central value $\rho_c$ of the rest-mass density of the
  NS, the ratio $r_p/r_e$ of the polar to equatorial radius of the NS,
  the baryonic ($M_b$) and gravitational ($M$) masses of the NS, the
  proper equatorial radius $R_e$, the ratio $J/M^2$ of total angular
  momentum and gravitational mass, and the compactness $C\equiv
  M/R_e$. The disk is characterized by the parameters $l_0$ and
  $\delta$ (see the text for description), its inner ($R_{\mathrm{in}}$) and
  outer ($R_{\mathrm{out}}$) radius, and its baryonic mass $M_{\rm torus}$.}
\end{center}
\end{table*}

Our Letter is organized as follows: in Section~\ref{id} we describe
the initial NS/torus configurations, and the numerical setup used to
study them. The dynamics of the collapse is reported in
Section~\ref{dynamics}, while the GW signal emitted during the AIC
event is described in Section~\ref{gws}, followed by a discussion
(Section~\ref{em}) of possible electromagnetic (EM) counterparts.  We
finally summarize and conclude in Section~\ref{conclusions}.

Throughout this Letter we use a spacelike signature of $(-,+,+,+)$ and
a system of units in which $c=G=M_\odot =1$ (unless specified
otherwise).

\section{Initial Data and Numerical Setup}
\label{id}

The initial data are built using the codes \texttt{RNS}
\citep{Stergioulas1995,Nozawa1998,Stergioulas2003} and \texttt{TORERO}
(\citealt{Corvinophd}). More specifically, the uniformly rotating NS
models are built using the numerical code \texttt{RNS}~with a
polytropic equation of state (EOS), $p=K \rho^\Gamma$, where $p$ and
$\rho$ are pressure and rest-mass density, respectively, while $K=100$
and $\Gamma=2$ are the polytropic constant and the polytropic
exponent, respectively. All the initial models are chosen to be on the
unstable branch~\citep{2011MNRAS.416L...1T} in order to be able to
compare the GW signal between a collapsing NS with and without a
torus. In the former case we assume that the torus has already
accreted enough mass on the NS to make it unstable, and we follow its
dynamics after that point. The NS collapse is triggered by reducing
its pressure by $0.1\%$ at the beginning of the simulations.

When we add an accretion disk to the rotating NS, we use the code
\texttt{TORERO}, which can compute stationary solutions of
axisymmetric disks around spinning BHs by solving the equations
described in~\cite{2004MNRAS.349..841D} and~\cite{Corvinophd}. In
these equations the self-gravity of the disk is neglected and hence
they are valid only when the mass of the disk is sufficiently small
compared to the mass of the central compact object.  We note that,
even if the torus is built assuming the spacetime of a spinning BH,
the violation of the Hamiltonian constraint introduced by having the
torus orbiting around a spinning NS is not large enough to affect the
dynamics of the system. This is due to the fact that the tori
considered here have masses which are a small fraction of the mass of
the NS. We have further verified that this violation decreases with
increasing resolution and that it is negligible compared to the
violations introduced during the evolution (i.e., the infinity norm of
the Hamiltonian constraint is similar during the collapse between the
models with and without torus). Furthermore, the initial violation
produced by adding the torus produces a spurious burst of radiation
that leaves the grid well before the beginning of the collapse and
hence it does not contaminate the GW signal computed from these
models. The tori are built by giving as input parameters to
\texttt{TORERO} the gravitational mass $M$ and the ratio of angular
momentum to gravitational mass, $J/M^2$, of the NS, and two parameters
$l_0$ and $\delta$ which define, respectively, the initial value of
the constant angular momentum in the disk and the depth level reached
by the disk inside the potential well. A negative value of $\delta$
represents a stable solution, while a positive value would give a
torus overflowing its Roche lobe.

Table~\ref{tab:id} lists the relevant properties of the NSs and tori
studied here.  We have chosen tori with masses $\sim 7\%$ of the mass
of the NS in order to have a non-negligible effect on the dynamics.

The evolution of these NS/torus configurations is computed using the
fully general relativistic magnetohydrodynamic code
\texttt{Whisky}~\citep{2005PhRvD..71b4035B,2007CQGra..24S.235G,giacomazzo11},
which adopts a flux-conservative formulation of the equations and
high-resolution shock-capturing schemes. All the results presented
here have been obtained by means of the Piecewise Parabolic Method,
while the Harten-Lax-van Leer-Einfeldt approximate Riemann solver has
been used to compute the
fluxes~\citep{2005PhRvD..71b4035B,2007CQGra..24S.235G}. All the
simulations were performed using an ideal-fluid EOS with
$\Gamma=2$. Magnetic field effects were not included in these
simulations and they will be the subject of future work. The evolution
of the spacetime was obtained using the \texttt{Ccatie} code, a
three-dimensional finite-differencing code providing the solution of a
conformal traceless formulation of the Einstein
equations~\citep{2007PhRvD..76l4002P}. Both the \texttt{Ccatie} and
\texttt{Whisky} codes benefit of the use of the Einstein
Toolkit~\citep{2012CQGra..29k5001L}.

The simulations were run using the vertex-centered fixed
mesh-refinement approach provided by the \texttt{Carpet}
driver~\citep{2004CQGra..21.1465S}.  We have used seven refinement
levels with the finest resolution being $0.1 M_{\odot}\approx
0.148\,\mathrm{km}$ and the coarsest resolution being $6.8
M_{\odot}\approx 10.04\,\mathrm{km}$. The finest grid has a radius of
$10 M_{\odot}\approx 14.77\,\mathrm{km}$ (which is sufficient to
completely cover the NS), whereas the coarsest grid extends to $998.4
M_{\odot}\approx 1474\,\mathrm{km}$. For all the simulations reported
here we have also used a reflection-symmetry condition across the
$z=0$ plane and a $\pi$-symmetry condition across the $x=0$
plane\footnote{Stated differently, we evolve only the region $\{x\geq
  0,\,z\geq 0\}$ and apply a $180^{\circ}$-rotational-symmetry
  boundary condition across the plane at $x=0$.}.

\begin{figure*}[p!]
  \begin{center}
    \begin{tabular}{cc}
      \includegraphics[width=0.4\textwidth]{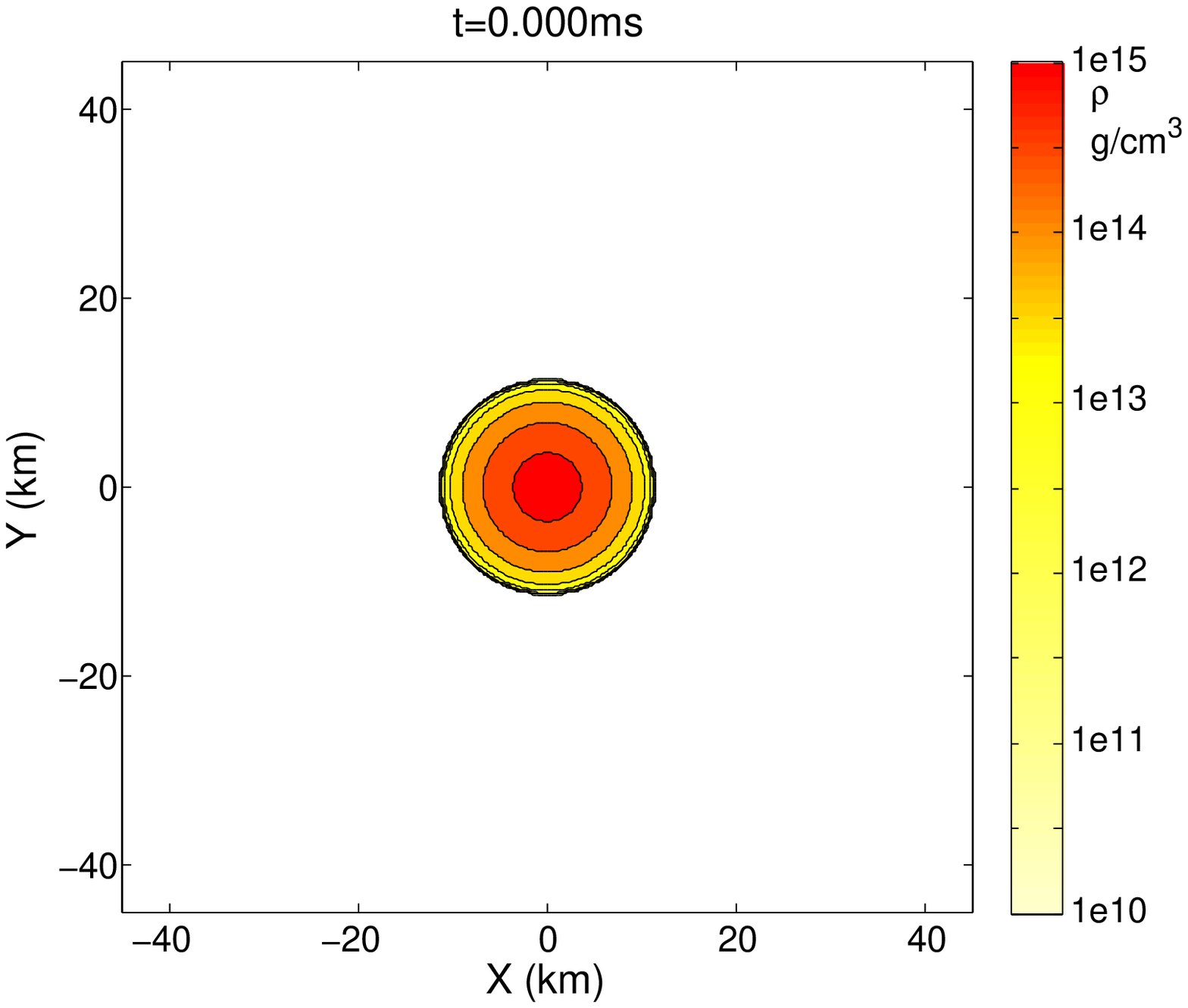}
      \includegraphics[width=0.4\textwidth]{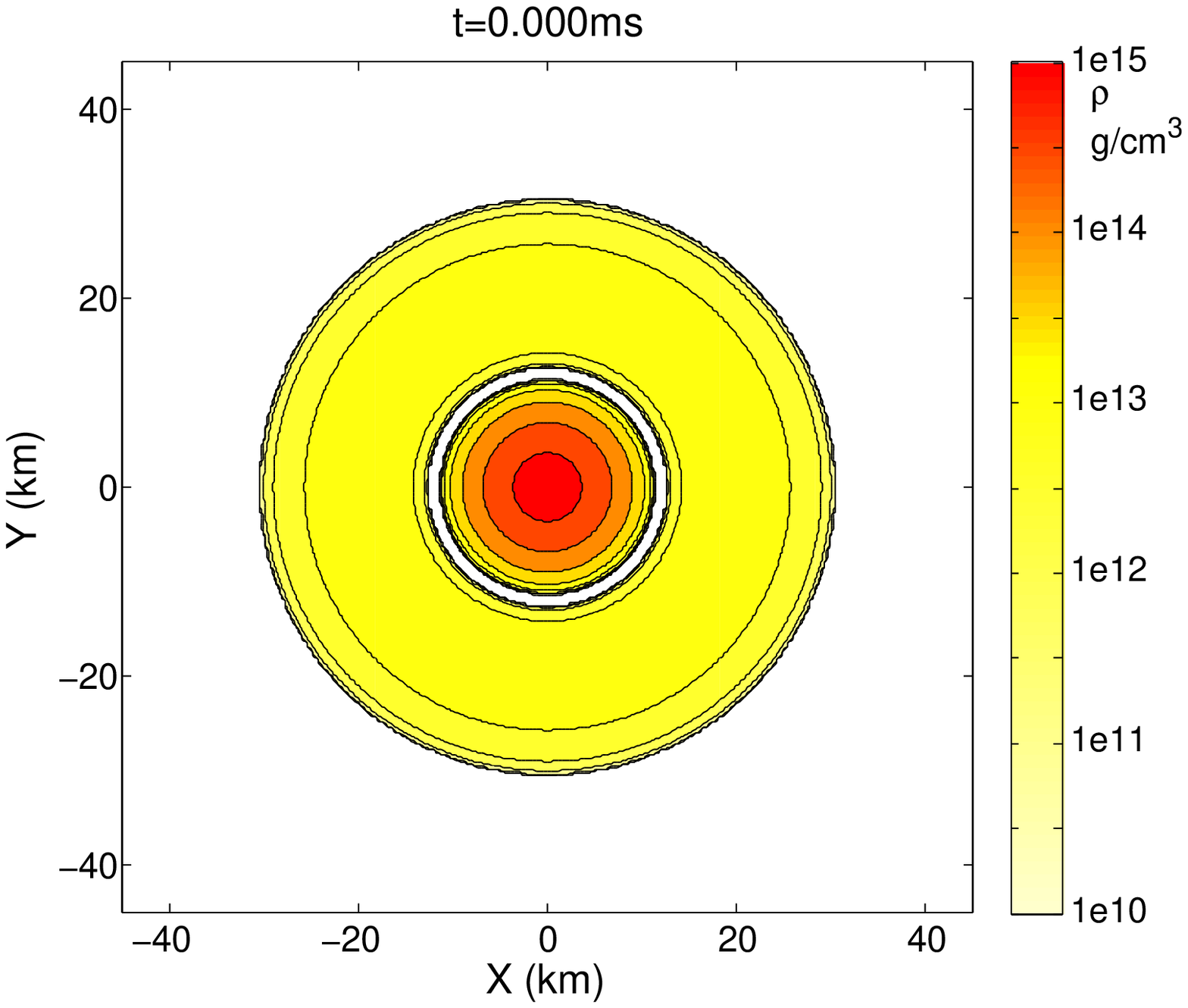}\\
      \includegraphics[width=0.4\textwidth]{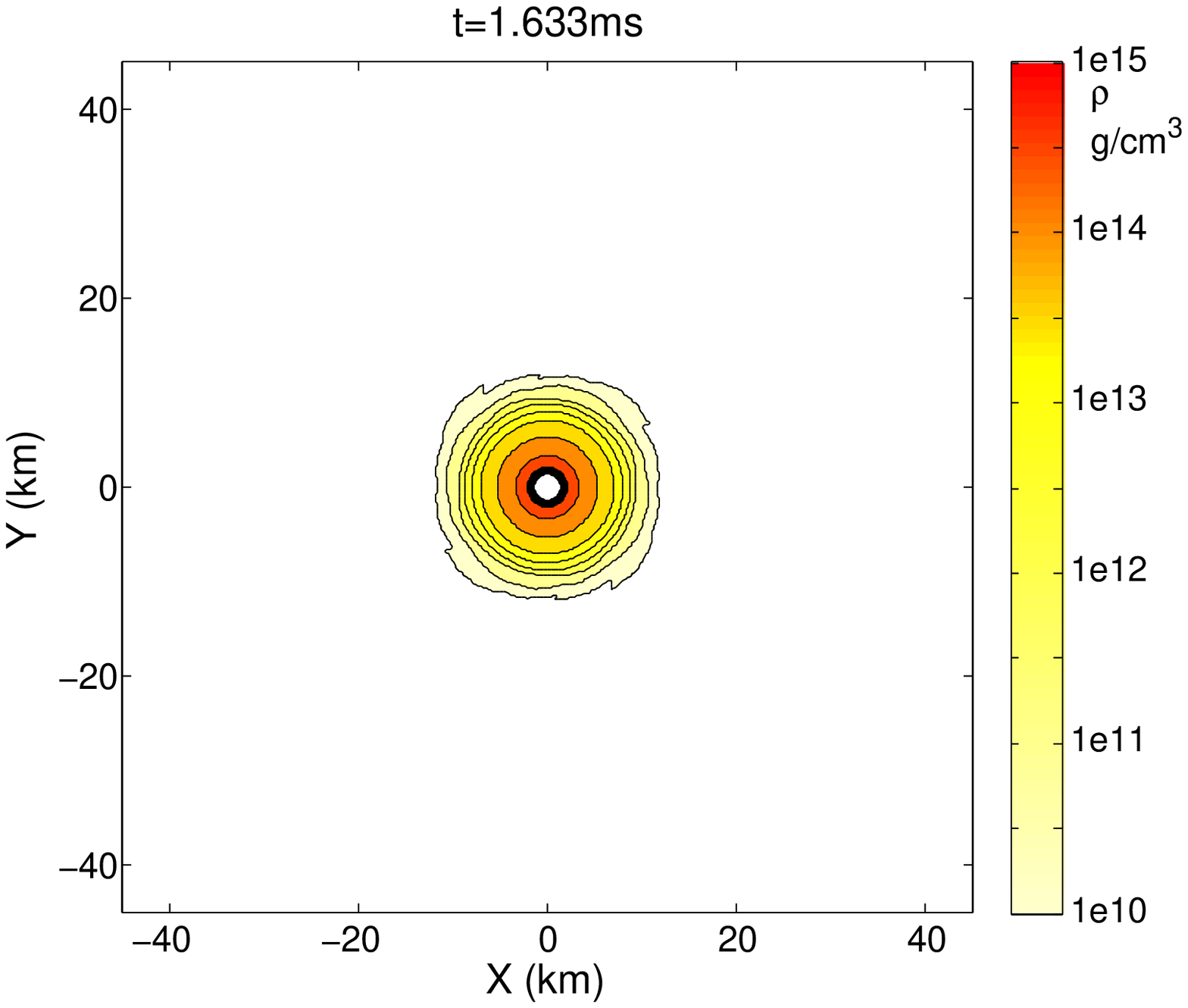}
      \includegraphics[width=0.4\textwidth]{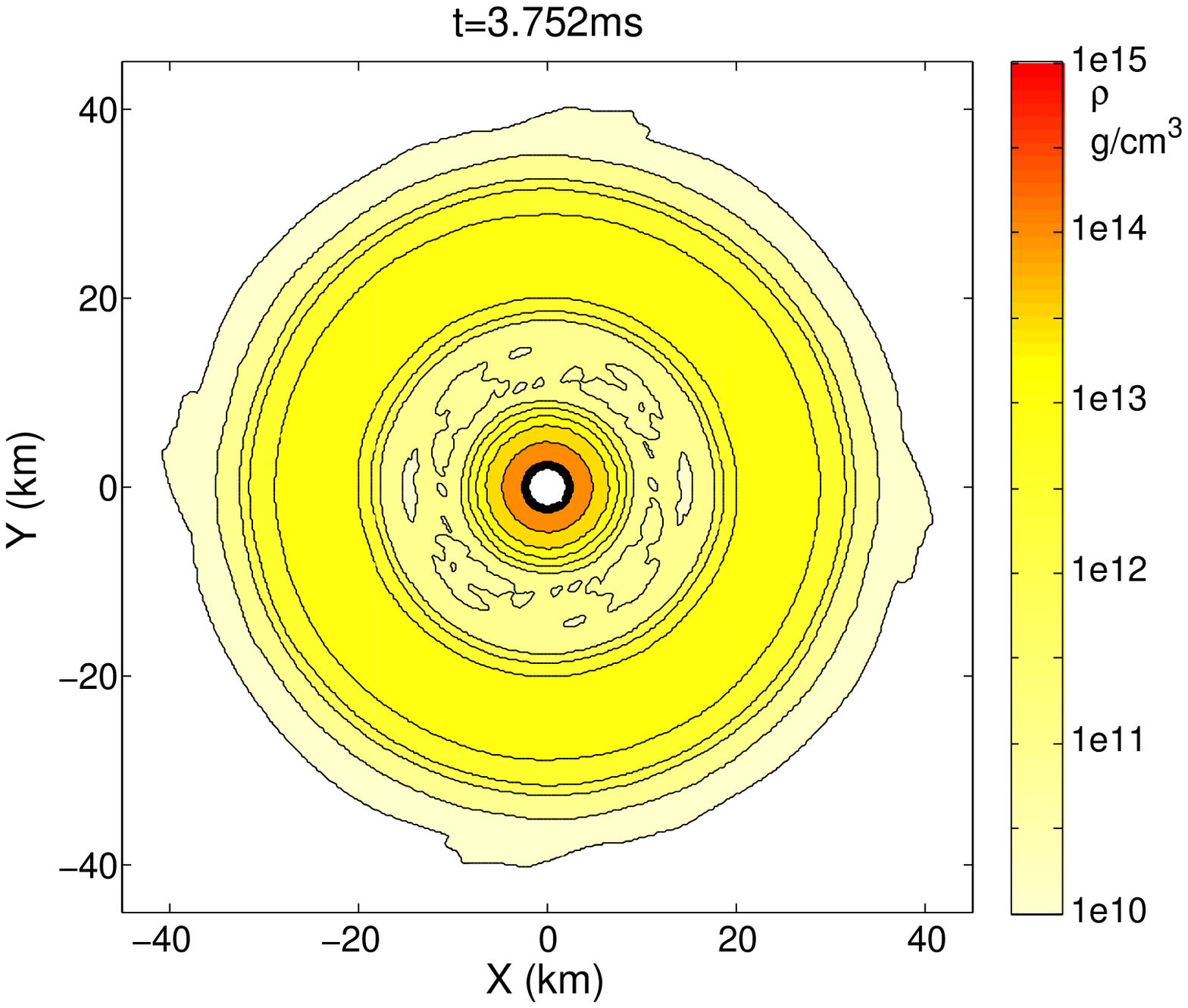}\\
      \includegraphics[width=0.4\textwidth]{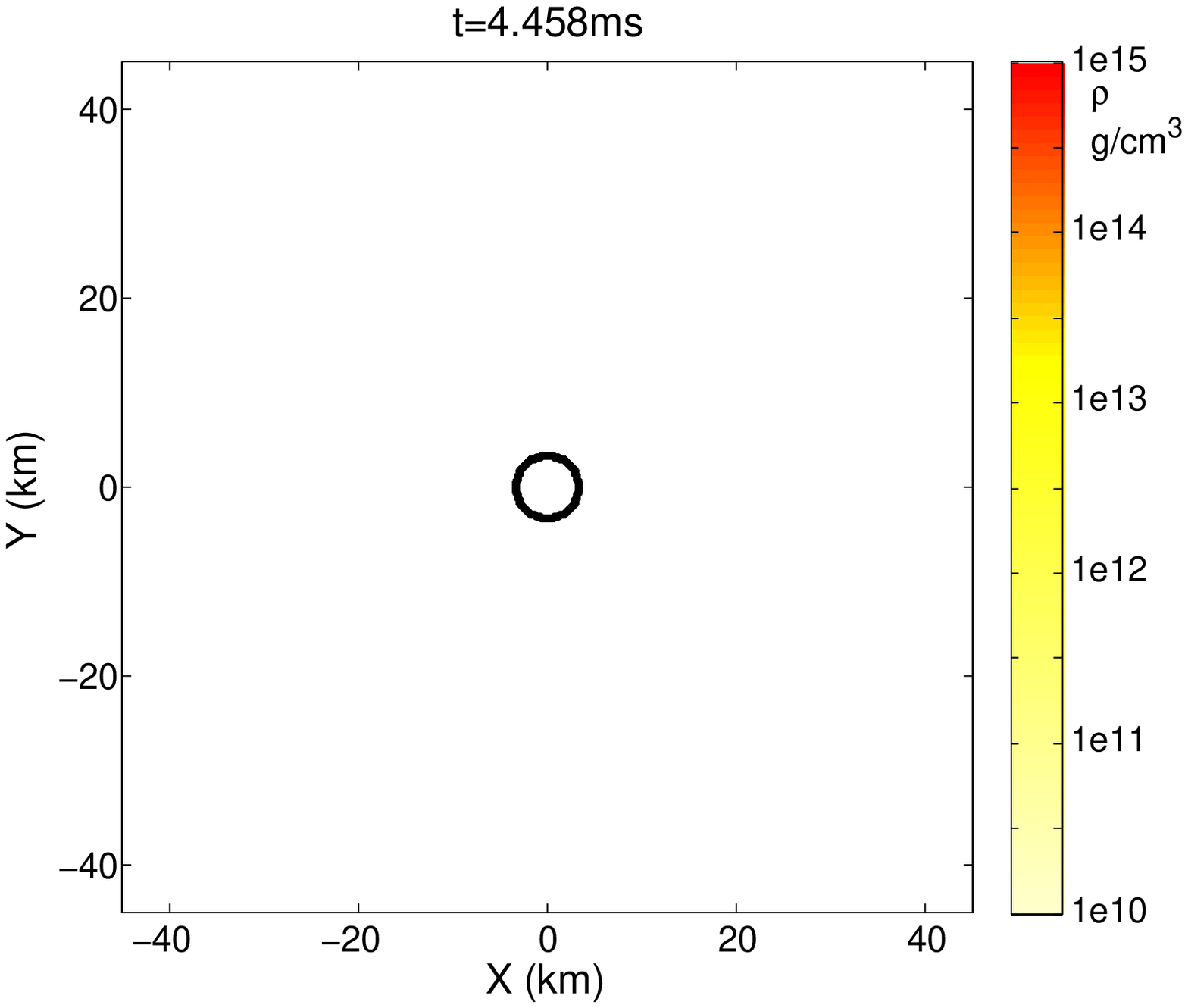}
      \includegraphics[width=0.4\textwidth]{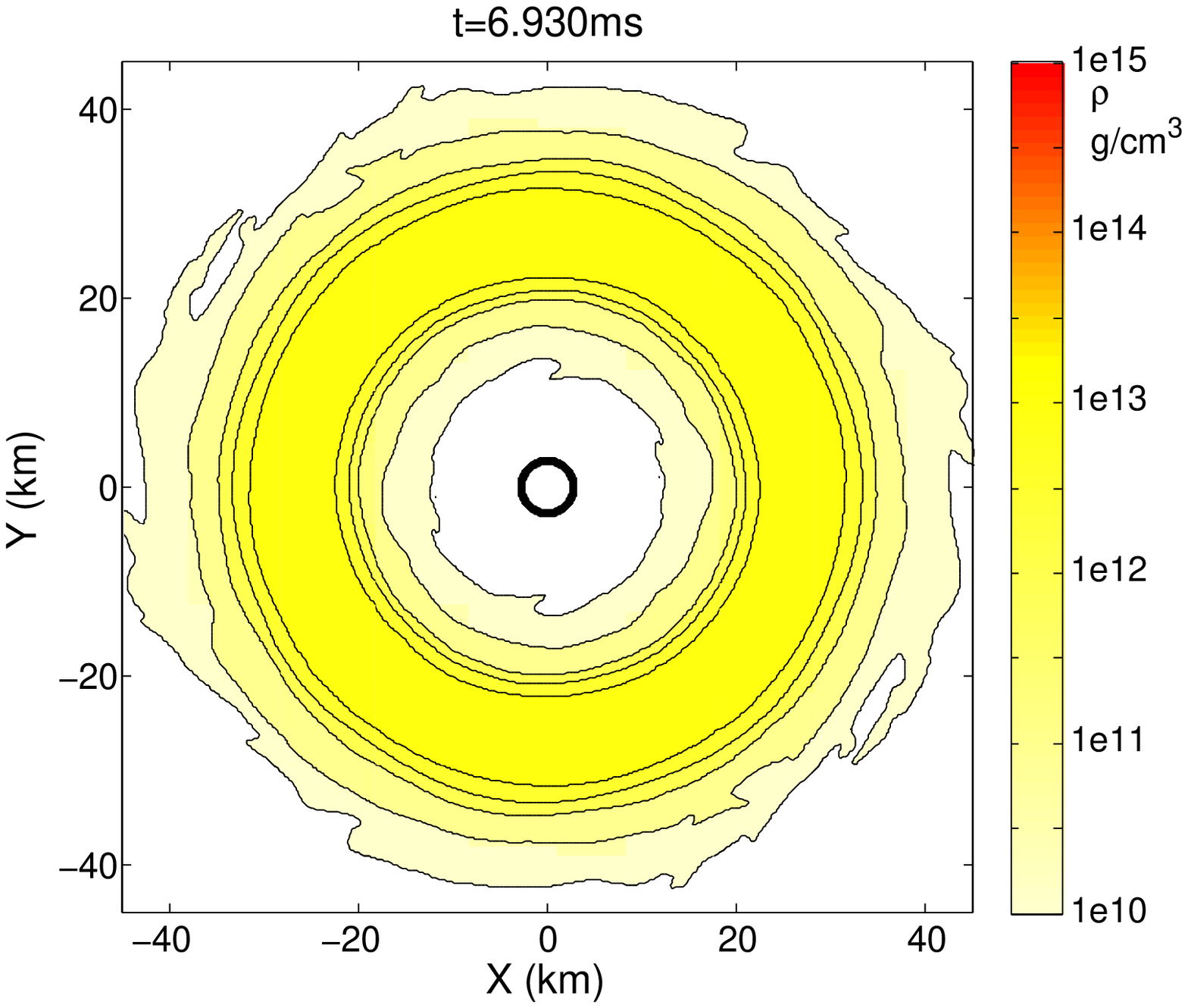}
    \end{tabular}
  \end{center}
  \caption{\label{fig:rho_xy}Evolution of the rest-mass density $\rho$
    on the equatorial plane for models \texttt{3a} (left panels) and
    \texttt{3b} (right panels). The top panels show the initial
    condition, the central one the moment of BH formation, and the
    bottom ones the end of the simulations. In the central and bottom
    panels, a thick black line denotes the position of the apparent
    horizon of the BH.}
\end{figure*}

\begin{figure*}[h!t!]
  \begin{center}
    \begin{tabular}{ccc}
      \includegraphics[width=0.3\textwidth]{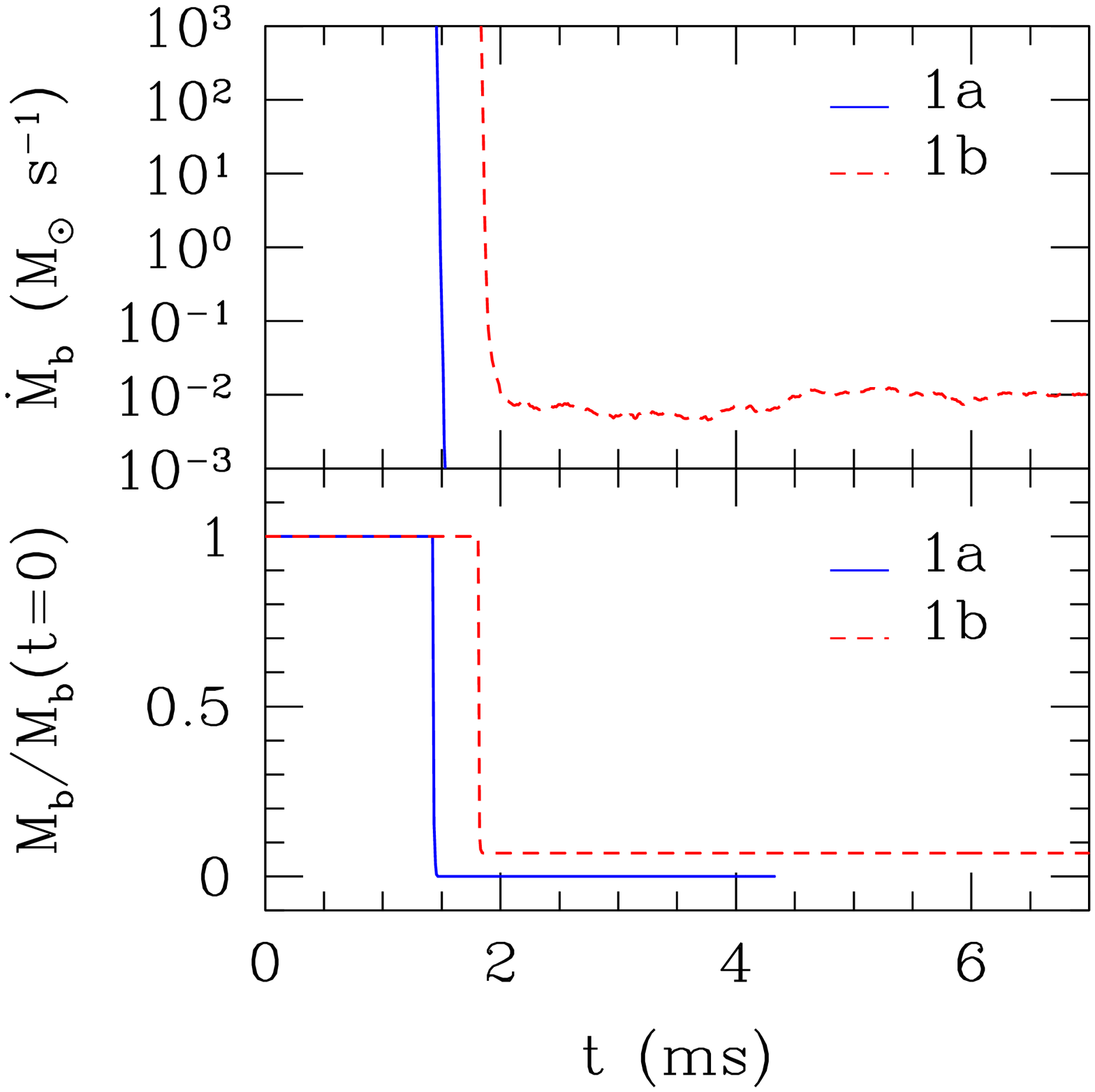}
      \includegraphics[width=0.3\textwidth]{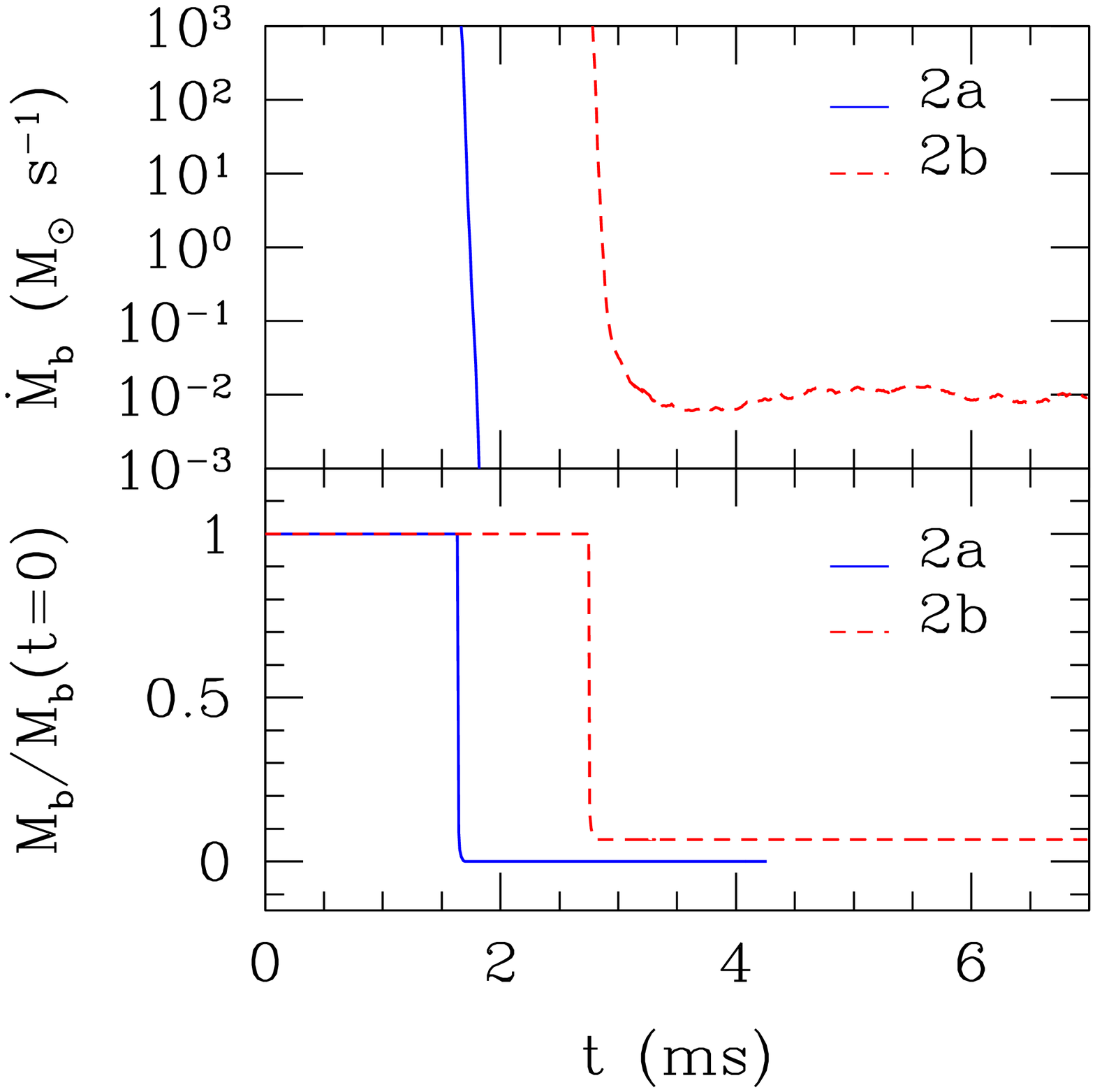}
      \includegraphics[width=0.3\textwidth]{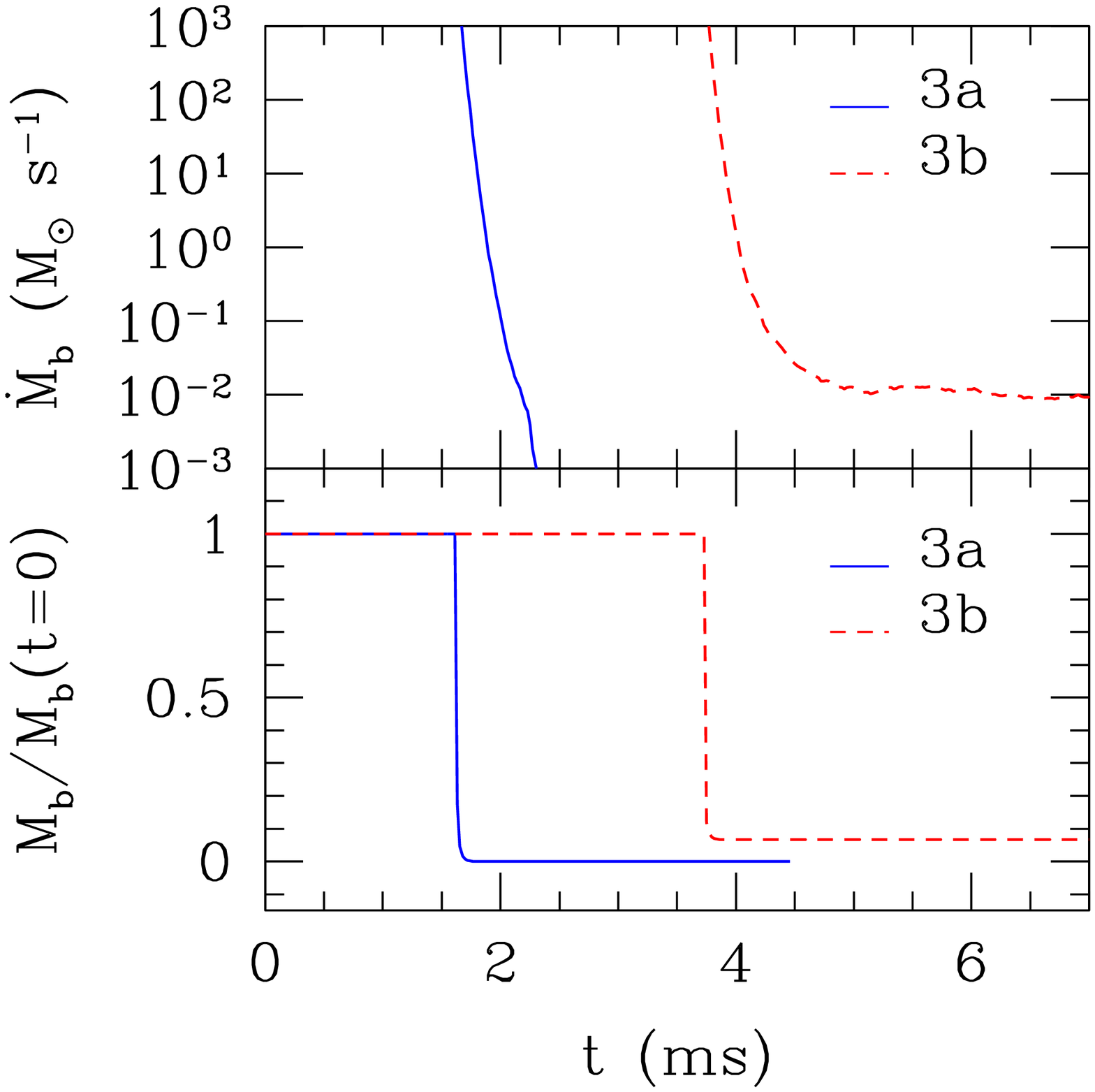}
    \end{tabular}
  \end{center}
  \caption{\label{fig:mass} Evolution of mass (bottom panels) and
    accretion rate (top panels) for models \texttt{1a} and
    \texttt{1b} (left panel), \texttt{2a} and \texttt{2b} (center
    panel), and \texttt{3a} and \texttt{3b} (right panel). In all the
    panels the blue solid line refers to the case without torus
    (\texttt{1a, 2a, 3a}) and the red dashed line to the case with the
    torus (\texttt{1b, 2b, 3b}).}
\end{figure*}

\section{Dynamics}
\label{dynamics}

Figure~\ref{fig:rho_xy} shows the evolution of the rest-mass density
$\rho$ on the equatorial $xy$ plane for models \texttt{3a} (left
panels) and \texttt{3b} (right panels). We remind the reader that the
only difference between the initial data of these two models is the
fact that we added a torus in the case of model \texttt{3b}. The top
panels show the initial data, the center ones the time of collapse to
BH, and the bottom panels the end of the simulations. While the
dynamics is essentially axisymmetric\footnote{In order to assess the
  impact of the use of $\pi$-symmetry on our results, we also
  performed some simulations without using it and we verified that the
  dynamics is not affected and that the evolution remains
  axisymmetric.} as it was already seen in the case of the collapse of
NSs without a disk~\citep{2007CQGra..24S.187B,2011PhRvD..84b4022G},
the main difference is that the torus resists the gravitational
collapse and continues to orbit and accrete onto the spinning BH
(bottom-right panel). Moreover, since the accreting torus triggers
oscillations in the NS density, the collapse is slightly delayed in
the case in which the NS is surrounded by a disk. This happens because
the NS oscillations prevent the central density from growing
exponentially from the very beginning, since each NS oscillation
produces an expansion and hence a small decrease of its central
density. This qualitative dynamics is similar also in the other cases
we studied (models 1a/1b and 2a/2b). In the bottom panels of
Figure~\ref{fig:mass} we show the evolution of the total baryonic mass
$M_b$, normalized to its initial value. All the cases which included a
torus at the beginning (red dashed lines) still preserve $\sim 7\%$ of
the initial mass outside the BH at the end of the simulations. On the
other hand, all the models without a torus (blue solid lines) do not
leave any mass outside the BH, as it was already seen in previous
simulations of the collapse of an NS to a BH in vacuum\footnote{Note
  that the presence of magnetic fields alters these
  conclusions. Simulations of magnetized, differentially rotating NSs
  ~\citep{duez1,duez2} show that, following the NS collapse, the BH is
  surrounded by a hot torus.}
~\citep{2007CQGra..24S.187B,2011PhRvD..84b4022G}. Another important
quantity is the accretion rate $\dot{M_{\rm b}}$ after the formation
of the BH (top panels of figure~\ref{fig:mass}). We compute this as
the integral of the flux of matter that crosses the apparent horizon.
In all the models that left a torus outside the BH, its early-time
value is found to be $\sim 10^{-2}~M_{\odot}~\mathrm{s}^{-1}$.

We note that in our simulations (as well as in previous
general-relativistic simulations of NS-NS mergers), the accretion is
triggered both by numerical viscosity and by the spacetime dynamics.
In order to assess the contamination from the numerical component
(i.e., the numerical viscosity), we performed simulations at different
resolutions, and found that the accretion rate remained at the same
level. Hence we can safely conclude that the contribution from
numerical dissipation is negligible in our results.  However note
that, in astrophysical disks, turbulent viscosity, likely generated by
the magneto-rotational instability \citep{BH}, causes angular momentum
transport and hence accretion. For an $\alpha$-viscosity parameter
$\sim 0.1$ \citep{SS}, we found that the viscous timescale in our
torus is $t_0\sim 0.1$~s, which would yield an even larger
post-collapse accretion rate $\dot{M}\sim M_{\rm torus}/t_0\sim
1~M_\sun$~s$^{-1}$.  Since this estimate is highly dependent on the
disk properties and its viscosity, we note that our computed
post-collapse $\dot{M_{\rm b}}$ should be rather considered as a lower
limit to what can be achieved by the system following an AIC event.

\begin{figure*}[h!t!]
  \begin{center}
    \begin{tabular}{ccc}
      \includegraphics[width=0.3\textwidth]{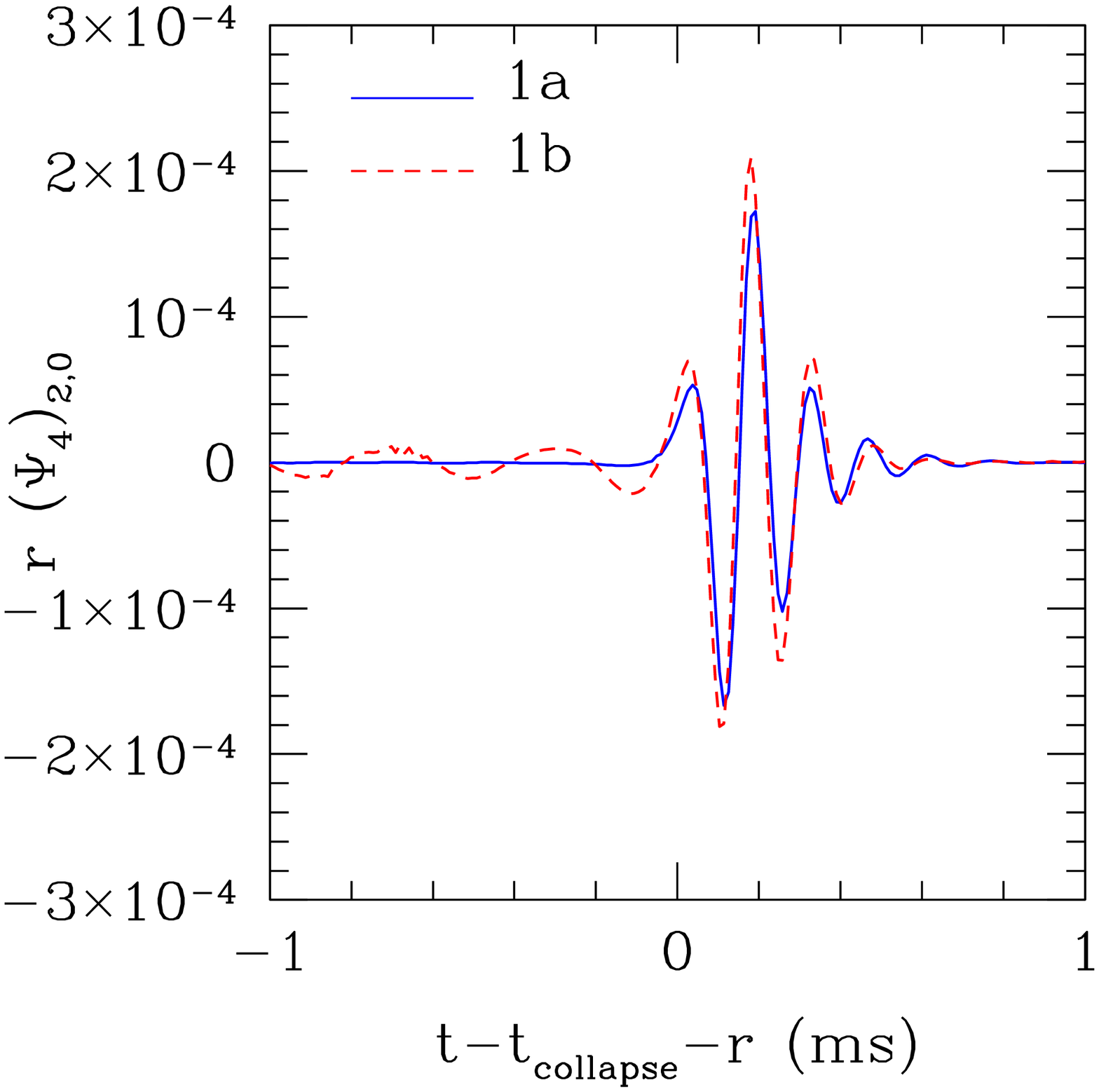}
      \includegraphics[width=0.3\textwidth]{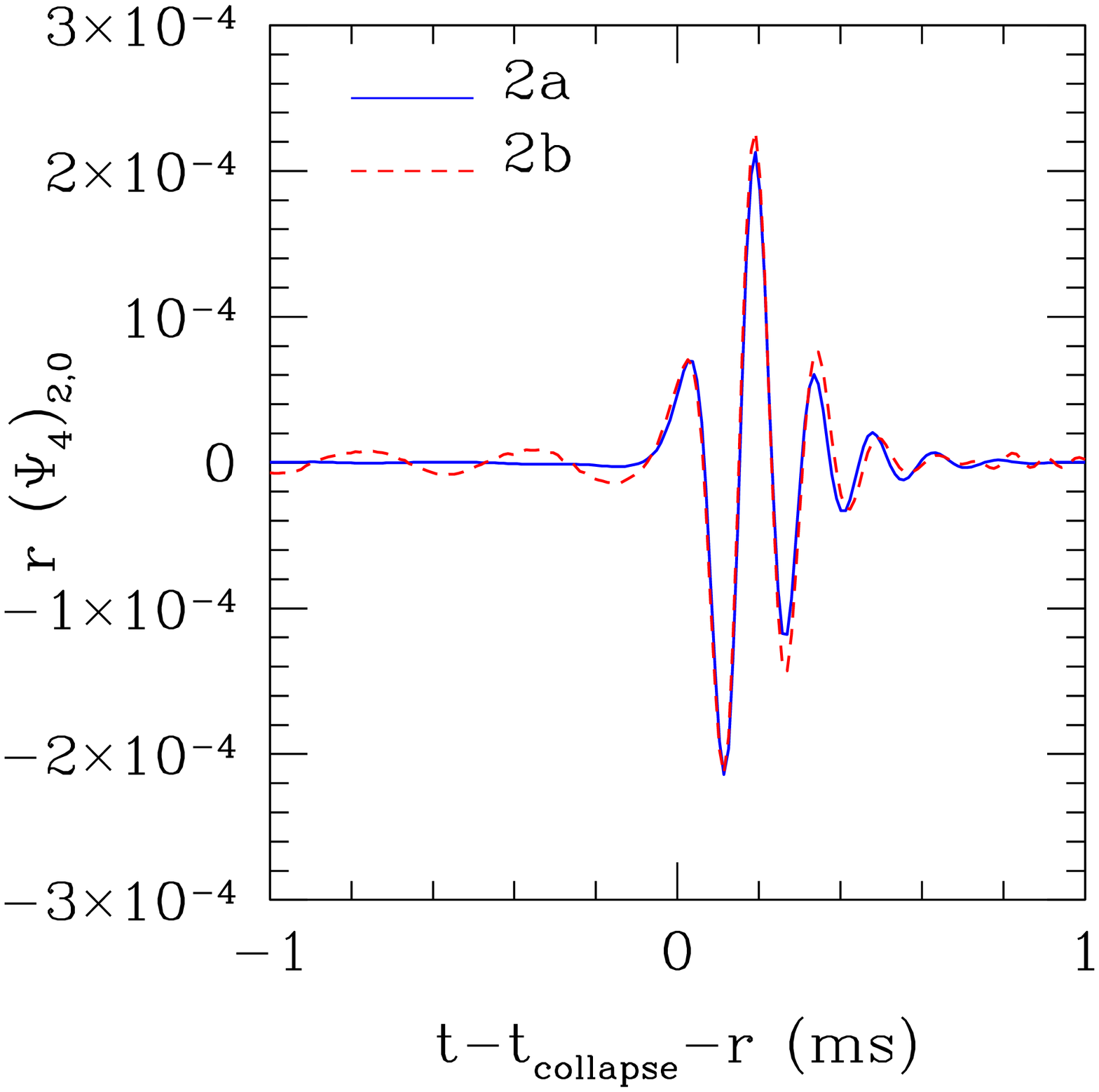}
      \includegraphics[width=0.3\textwidth]{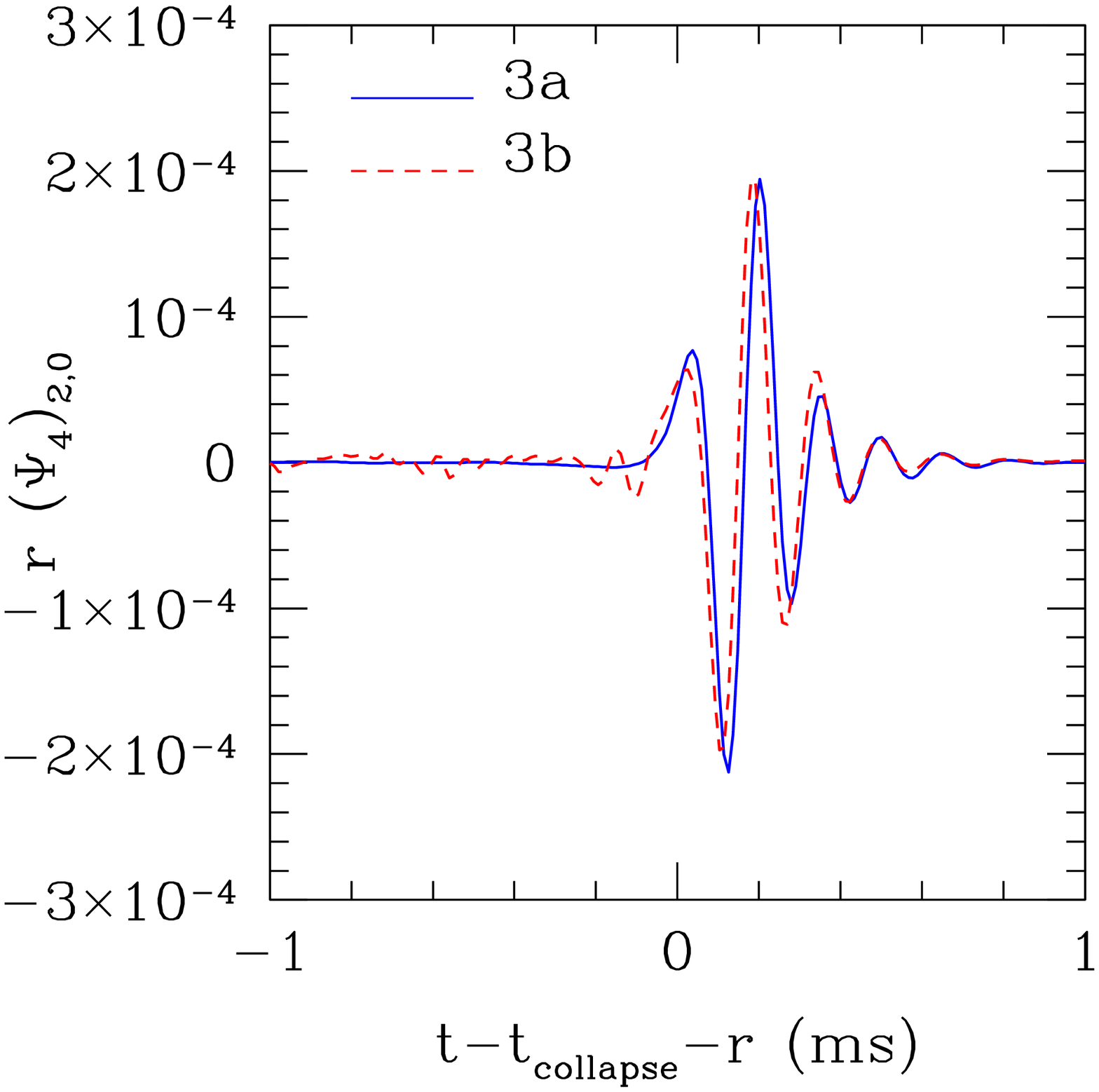}
    \end{tabular}
  \end{center}
  \caption{\label{fig:psi4} Gravitational wave signal $r
    (\Psi_4)_{(l=2,m=0)}$ for models \texttt{1a} and \texttt{1b} (left
    panel), \texttt{2a} and \texttt{2b} (center panel), and \texttt{3a}
    and \texttt{3b} (right panel). All the signals are extracted at
    $r=100 M_{\odot}\approx 148\,\mathrm{km}$. In all the panels the
    blue solid line refers to the case without torus (\texttt{1a, 2a,
      3a}) and the red dashed line to the case with the torus
    (\texttt{1b, 2b, 3b}). The time is shifted by the time of collapse
    $t_{\mathrm{collapse}}$ (defined as the moment at which an apparent horizon
    is formed).}
\end{figure*}

\section{Gravitational Waves}
\label{gws}

Figure~\ref{fig:psi4} displays the GW signal emitted by the AIC of an
NS to a BH in our six models, and in particular the dominant $l=2,m=0$
mode computed using the Weyl scalar
$\Psi_4$~\citep{baker2002,2012CQGra..29k5001L}. The blue solid lines
refer to the cases without a torus, while the red dashed lines show
the signal emitted when the torus is present. Since the collapse is
essentially axisymmetric, $l=2,m=0$ is the dominant mode, and the
signal is given by an exponential increase followed by the ring-down
of the BH. No significant difference can be seen in the signal between
the models with and without a torus, except for the fact that the GWs
emitted by the models with a torus show some small oscillations before
the collapse. These are related to the NS oscillations triggered by
the accretion of the disk, as previously mentioned in
Section~\ref{dynamics}. Similarly to the case of the collapse of an NS
in vacuum, the GW emission would be detectable by advanced LIGO/Virgo
only if the sources were located in our Galaxy. On the other hand,
with the Einstein Telescope, the signal could be detected for sources
at distances of up to $\sim 1$ Mpc.\\

\section{Electromagnetic Emission}
\label{em}

As described in Section~\ref{dynamics}, the presence of a rapidly
accreting torus left around the newly born BH is a distinctive feature
of the AIC, differentiating it from the direct collapse of an NS into a
BH, which leaves no mass for the BH to accrete from.

The accretion rates that our AIC events predict
(cf. Figure~\ref{fig:mass}), on the order of $\sim 10^{-2}M_{\odot}~
\mathrm{s}^{-1}$, are within an order of magnitude of those obtained in the
merger of two NSs (e.g., \citealt{rez10}). For a typical mass-to-energy
conversion efficiency $\sim 0.1$, these AICs have luminosities
sufficiently high to constitute enticing candidates for short GRBs.
In fact, assuming an energy-to-$\gamma$-ray conversion efficiency
$\sim$ several tens of percent (as expected for short GRBs, e.g.,
\citealt{zhang07}), our simulated AIC events would have peak
luminosities in $\gamma$-rays on the order of $\sim
10^{51}$~erg~s$^{-1}$, compatible with at least a fraction of short
GRBs.  Clearly, the predicted accretion rates are dependent on the
torus mass, and a more comprehensive exploration will be performed in
future work.  

Short GRBs from the AIC of an NS in a binary are envisioned to have
the observer located out of the binary
plane~\citep{2005astro.ph.10192M}, and a jet launched along the NS
rotation axis -- either as the result of $\nu\bar{\nu}$ annihilation
(e.g., \citealt{aloy05}) or of magnetohydrodynamic effects (e.g.,
\citealt{mhd}); furthermore, they are expected to be more often found
in early-type or globular cluster environments
\citep{2006ApJ...643L..13D}. A detailed computation of their expected
cosmological rates, and a comparison with the observed population of
short GRBs, is reserved for future work. In particular, we will
examine the dependence of the AIC accretion rates on the disk mass,
and hence assess the likelihood of their occurrence in X-ray binaries
(where disk masses are smaller than the ones considered here) and
fallback disks around isolated NSs (e.g., \citealt{Perna2000} and
references therein). Finally, within the connection short GRBs/AIC
events, we further note that the observed X-ray
flares~\citep{margutti} and extended emissions~\citep{norris}, lasting
up to several hundreds of seconds, cannot be directly explained as the
result of activity from the central engine (which in an AIC is much
shorter than the flare duration), unless the disk undergoes some
instability (e.g. \citealt{paz}; \citealt{pz}).

\section{Conclusions}
\label{conclusions}

We have performed the first three-dimensional, fully general
relativistic simulation of the AIC of an NS to a BH; we have examined
the GW signal expected from such events, and computed the accretion
rates onto the newly formed BH.  We find that the main component
($l=2, m=0$ mode) of the GW signal is very similar to the one from the
collapse of an NS into a BH in vacuum. However, and most importantly,
while the direct collapse of a non-magnetized NS onto a BH leaves no
mass behind (hence making unlikely the presence of any EM detectable
counterpart), an AIC is followed by rapid accretion of the disk onto
the newly formed BH.  Our computed accretion rates, comparable to
those obtained in simulations of NS-NS mergers, make AIC events
potential engines for short GRBs. NSs in X-ray binaries (e.g.,
\citealt{bildsten97}), isolated NSs accreting from fallback
disks~\citep{chev}, and NSs produced by the AIC of a white
dwarf~\citep{2010PhRvD..81d4012A} are astrophysical scenarios which
might host the AIC of an NS onto a BH.

\acknowledgments We thank Giovanni Corvino for providing us with the
code \texttt{TORERO} and Roland Haas for assistance in computing the
accretion rates. We also thank Eleonora Troja, Cole Miller, and the
anonymous referee for carefully reading the manuscript and for their
useful comments. Numerical simulations were performed on the cluster
RANGER at the Texas Advanced Computing Center (TACC) at The University
of Texas at Austin through XSEDE grant No. TG-PHY110027. B.G. and R.P.
acknowledge support from NSF grant No. AST 1009396.

\bibliographystyle{apj}

\end{document}